\documentclass[twocolumn,showpacs,preprintnumbers,amsmath,amssymb]{revtex4}

\usepackage{graphicx}
\usepackage{dcolumn}
\usepackage{bm}

\begin{document}

\preprint{}

\title{Large-Mass Ultra-Low Noise Germanium Detectors:\\ Performance and 
Applications in Neutrino \\and Astroparticle Physics}

\author{P.S. Barbeau}
\author{J.I. Collar}%
 \altaffiliation[contact author: ]{collar@uchicago.edu}
\affiliation{%
Department of Physics,  Enrico 
Fermi Institute and Kavli Institute for Cosmological Physics, 
        University of Chicago, Chicago, IL 60637, USA\\
}%
\author{O. Tench}
\affiliation{
CANBERRA Industries, 
         800 Research Parkway, Meriden, CT 06450, USA\\
}%

\date{\today}

\begin{abstract}
A new type of radiation detector, a p-type modified electrode 
germanium diode, is presented. The prototype 
displays, for the first time, a combination of features 
(mass, energy threshold and background expectation) required for a 
measurement of coherent neutrino-nucleus scattering in a nuclear reactor 
experiment. The device hybridizes the mass and energy resolution of a 
conventional HPGe coaxial gamma spectrometer with the low electronic 
noise and threshold of a small x-ray semiconductor detector, also 
displaying an intrinsic ability to distinguish multiple from 
single-site particle interactions. The present performance of the prototype and possible further 
improvements are discussed, as well as other applications for this new 
type of device in neutrino and astroparticle physics (double-beta 
decay, neutrino magnetic moment and WIMP searches).
\end{abstract}

\maketitle

\section{INTRODUCTION}

From an experimental point of view, the detection of the soft 
nuclear recoils expected from coherent scattering of low-energy 
neutrinos off nuclei \cite{freedman,drukier} remains a daunting 
challenge. 
The applications of a capable detector to fundamental and 
 applied physics  would be numerous: a large coherent 
 enhancement to the cross section for this process results in expected 
recoil rates in the hundreds 
 per kg of target per day from reactor antineutrinos, affording a radical 
 detector compaction. Of all man-made neutrino sources, power reactors 
 offer by far the largest flux, with essentially all (anti)neutrino 
 energies being able to interact coherently. The difficulty resides 
 in a majority of these 
 recoils falling well below the energy threshold of conventional 
 detector technologies. Until now, no device 
has met the minimum target mass ($\underset{\sim}{>}1$ kg) and energy threshold 
($E_{rec}\underset{\sim}{<}1$ keV) required for such a 
measurement, even though unrealized proposals abound 
\cite{drukier,ieee,blas}. Not only the recoil energies themselves are low, 
but only a fraction of these are generally available in a readily detectable 
form (ionization, scintillation). A characterization of this 
fraction (the ``quenching factor'') and a precise definition of signal 
acceptance near the detector threshold are necessary first steps in the 
development of a coherent neutrino detector. In a companion paper 
\cite{other}, an 
approach to this challenging in itself calibration exercise has been 
developed.

The advent of a coherent neutrino detection technology would enable true technological applications 
\cite{leocastle}, for instance, non-intrusive monitoring of nuclear reactors against illegitimate uses 
(e.g., fuel rod diversion, production of weapon-grade material)
with a truly compact device, improving on proposed methods \cite{bernstein1}
that rely on standard (charged current) ton-sized neutrino detectors. Beyond 
these, the interest in observing this process is not merely academic: a 
neutral-current detector responds essentially the same way to all 
known neutrino types \cite{all}.
Therefore, the observation of neutrino oscillations in such a device 
would constitute as close to 
 {\it direct} evidence for sterile neutrino(s) as is possible. 
Separately, the cross section for this process is critically dependent on 
a neutrino magnetic moment: disagreement with the Standard Model prediction 
can reveal a finite value 
presently beyond reach \cite{dodd}. Recently the strong sensitivity of the process to non-standard 
neutrino interactions with quarks \cite{new} and the effective neutrino 
charge radius \cite{bernabeu} has been emphasized. Other fundamental physics 
applications can be listed: a precise enough measurement 
of the cross section
would constitute a sensitive probe of the weak nuclear charge
\cite{larry}.
Coherence plays a most important role in neutrino 
dynamics within supernovae and neutron stars \cite{freedman}, adding to the 
attraction of a laboratory measurement. 
In particular, a measurement 
of the total (flavor-independent) neutrino flux from a nearby 
supernova using a massive enough coherent detector would be of capital importance to 
help clarify the exact oscillation pattern followed by the neutrinos 
in their way to the Earth \cite{beacom}.

\section{A FIRST COHERENT NEUTRINO DETECTOR}

The Low-Background Detector Development group at the Enrico Fermi Institute 
has investigated several new technologies, each in principle 
capable of meeting the three goals (energy threshold, background and 
minimum detector mass) required for a successful measurement of this mode of neutrino 
interaction in a power reactor. In this paper we concentrate on what 
is 
presently considered the 
most promising path towards this measurement. The prototype described 
here
exhibits an active mass of 475 g, sufficient for a measurement of the cross 
section. The technology is however readily scalable to a mass O(10) kg, 
necessary 
for most of the applications mentioned 
above, in the form of a small array of detectors.

While an average of 20-100 eV in deposited ionization energy is required 
to generate a signal carrier (electron, photon) in proportional 
chambers or 
scintillating particle detectors, semiconductors offer the same for a 
$\sim$3 eV modicum. It would then make sense to examine the 
possibility of using large intrinsic germanium detectors for this application. 
Unfortunately, the large capacitance of a germanium diode of this 
mass ($\sim$1 kg) 
results in prohibitive electronic noise levels, i.e., ionization 
energy thresholds of a few keV. When a low-energy  quenching factor 
for Ge recoils 
of O(20)\%
\cite{ge}
is taken into consideration, the situation is seen to be hopeless for 
any man-made low-energy neutrino source. The TEXONO 
collaboration has recently proposed \cite {texono} to bypass this 
impediment by using arrays of 
commercially-available smaller n-type germanium 
diodes (5 g) where the low capacitance ($\sim$ 1 pF) results in 
$\sim$300 eV hardware ionization thresholds. In those conditions, as discussed below, 
coherent signal rates of O(10) detectable recoils kg$^{-1}$ day$^{-1}$ are 
possible from power reactors. 
The drawbacks of that approach are evident: a single kg of target 
mass requires a multitude of amplification channels and a large 
monetary investment. Near threshold, a much higher rate of 
unrejectable electronic noise-induced events per unit mass would be sustained 
than in a single-channel 1 kg crystal of comparable 
threshold, if such a thing could be built. Finally, the 
unfavorable peak-to-Compton 
ratio for  detectors this small results in partial energy depositions from environmental 
radioactive backgrounds being 
strongly favored, swamping the signal region \cite {texono}. 

Large germanium detectors with $<$1 pF capacitance and $\sim$1 kg of mass 
can be built, 
and this is the approach taken here. Modifications of the usual 
electrode structure in a variety of detector geometries 
\cite{luke1,luke2} can lead to these uncommonly low values of the capacitance, 
often at the expense of degraded energy resolutions, which are 
evidently not the biggest issue in the present application (the 
intrinsically-low resolution for small energy depositions being more important). An
800 g modified coaxial detector with a n$^{+}$ central electrode 
(n-type)
just 1 mm in radius and 0.5 mm deep achieved a 270 eV FWHM 
noise level \cite{luke2}, a sizeable improvement over 
a standard coaxial detector of the same mass. 
The measured capacitance (1 pF) was in excellent agreement with 
expectations based on the chosen electrode geometry \cite{luke2}. 
In order for the carriers to be efficiently collected 
along the axis of such a detector a gradient of impurities in the 
crystal must be present for space-charge effects to establish 
an effective electric field in this direction.  Material with these
characteristics is actually commonly present in germanium ingots as a 
result of segregation processes during crystal growth \cite{ingot}. 
The detector in \cite{luke2} was inspired by searches for the 
``Cosmion'' \cite{cosmion}, a low-mass particle dark matter candidate thought at the 
time to be able to explain the now defunct ``solar neutrino problem''. Due to 
their light mass ($\sim$ 1 GeV) these particles were expected to 
produce relatively soft recoils. Cosmions 
were experimentally ruled out soon after and no further  
progress nor interest in these detectors
appears in the literature.

A noise level of 270 eV (resulting in a threshold two to three times 
higher) would still be insufficient for a reactor 
 experiment looking for  coherent neutrino scattering. However, a number of 
improvements could in principle reduce this to a threshold $\sim$100 
eV or even 
lower in such a large detector, which would yield a comfortable signal 
rate for present purposes. Since 
a large fraction
of the noise figure in the seminal work of \cite{luke2} was due to the electronic 
characteristics of the field effect transistors (JFETs) available 20 years 
ago, it seemed timely to reconsider this approach in the light of the most recent technology. 
A number of possible improvements were envisioned:\\

$\bullet$ Replacing the resistor-feedback preamplifier used in 
\cite{luke2} by a lower noise optical- or transistor-reset type is observed to  
reduce on its own the electronic noise level by a factor $\sim$2 \cite{pulsed} in
detectors with a capacitance  $C_{D}\sim$1 pF using the same JFET as in 
\cite{luke2} (Texas Instruments 2N4416). This is at the expense of 
a reduced energy throughput \cite{pulsed}, 
which is however not a concern in low-counting rate experiments. \\

$\bullet$ Great progress in reducing JFET capacitance ($C_{F}$) and noise 
voltage  ($V_{n}$) has taken place since \cite{luke2}. The 2N4416 JFET is now 
fully outdated \cite{outdate} ($C_{F}$=4.2 pF, $V_{n}$= 2 nV/$\sqrt{Hz}$ 
at 295 $^{\circ}$K and 10 
kHz). A state-of-the-art  EuriFET ER105 
($C_{F}$=0.9 pF, $V_{n}$= 1.6 nV/$\sqrt{Hz}$ 
at 295 $^{\circ}$K and 10 
kHz), proprietary to CANBERRA Industries, provides a perfect capacitance match to a 
$C_{D}\sim$ 1 pF and in the process the theoretically best possible signal-to-noise ratio 
\cite{match,radeka}. The electronic noise level (as described 
by the energy resolution using a pulser) that can be expected 
from these technological improvements is described for Germanium by 
the expression \cite{radeka,pentafet}:

\begin{equation}
FWHM=(41 ~eV) ~V_{n} (C_{F}+C_{D})/\sqrt{\Delta t},
\end{equation}

 with all units as  above and $\Delta t (\mu$s) being the shaping 
 (integration) time of the (second 
stage)
spectroscopy amplifier. For $\Delta t = 8 \mu$s as in \cite{luke2}, 
an optimal $\sim$45 eV FWHM is to be expected, in contrast to 
$\sim$130 eV for the 2N4416 (this last figure in fair agreement with the observations 
in \cite{luke2} after introducing the factor $\sim$2 noise increase from
the resistor-feedback preamplifier). This noise figure is precisely at 
the level of the very best performance presently achievable with 
small $C_{D}$=1 pF, low-$C_{F}$, low-$V_{n}$ silicon x-ray detectors 
\cite{outdate,pentafet}:
the conclusion is that, in principle, it could
also be achieved with a modified-electrode 
large-mass HPGe 
detector. These
promising expectations may nevertheless be somewhat limited by the effect of 
lossy dielectrics \cite{lossy} and leakage currents \cite{leak}, which 
are seen to start to dominate 
noise contributions precisely at the scale of 50 eV FWHM (see 
discussion below). It is 
nevertheless worth mentioning that JFET technology continues to 
improve, and that experimental $C_{F}$=0.4 pF proprietary JFETs are available to 
CANBERRA Industries. This represents a full order of magnitude improvement 
since \cite{luke2}.\\

\begin{figure}
\includegraphics[width=9cm]{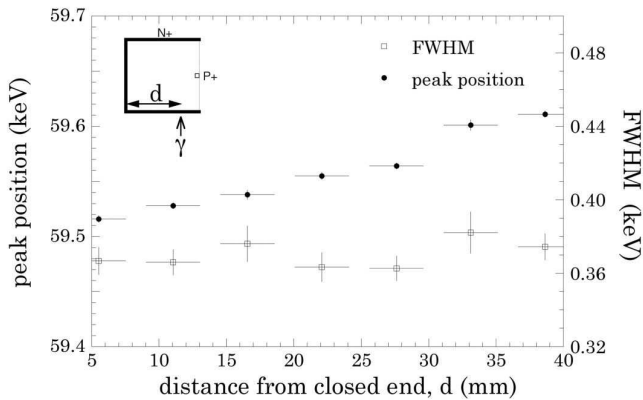}
\caption{\label{fig:epsart}Energy resolution (FWHM) and effective gain shift observed 
using low-energy gamma emissions from a collimated $^{241}$Am source at different 
positions along the longitudinal HPGe crystal axis. The tiny axial 
dependence of the second 
(0.15\% maximum variation)
demonstrates an optimal charge collection even in the presence of a 
modified electrode configuration.}
\end{figure}

\begin{figure}
\includegraphics[width=9cm]{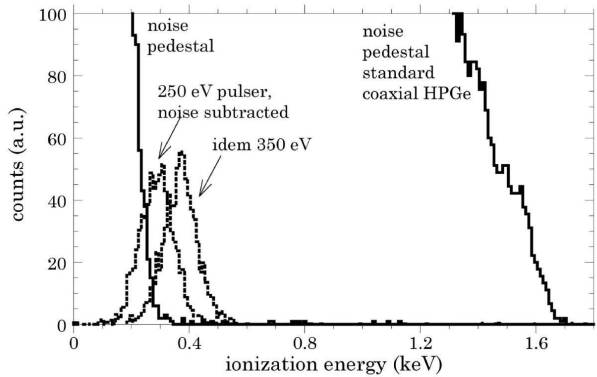}
\caption{\label{fig:epsart}A comparison of the energy threshold ($\sim$ 
330 eV, 5 sigma from noise centroid) in the 
modified electrode HPGe with that of a conventional coaxial detector of the 
same mass ($\sim 475$ g), typically in the few keV region (the particular 
one used for the figure being relatively low in noise). No 
instabilities in the threshold have been observed 
in five months of continuous 
detector operation. Energies are electron-equivalent, i.e., ionization.}
\end{figure}

\begin{figure}
\includegraphics[width=9cm]{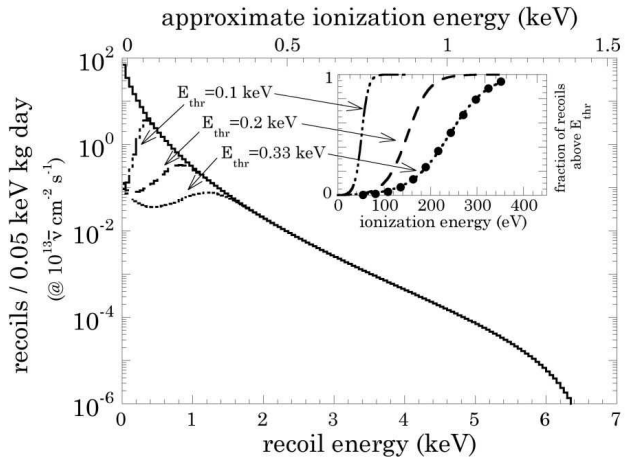}
\caption{\label{fig:epsart}Effect of detector threshold and energy 
resolution in the differential rate of recoils expected from reactor 
antineutrinos (see text).}
\end{figure}

\begin{figure}
\includegraphics[width=9cm]{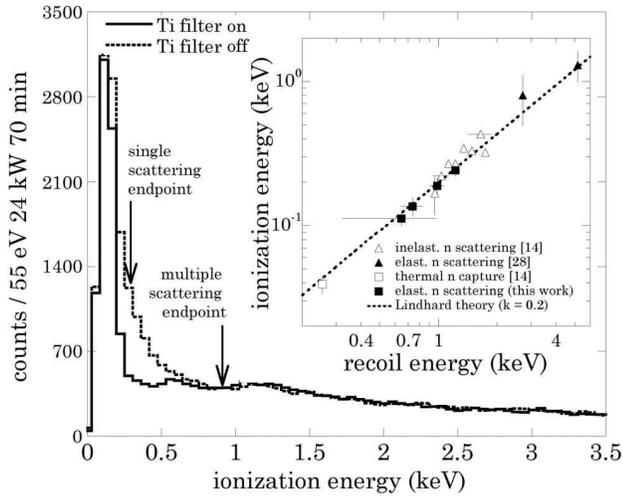}
\caption{\label{fig:epsart}Exposure of the prototype to a 
monochromatic 24 keV reactor neutron beam custom-built to mimic reactor 
antineutrino recoils \protect\cite{other}. A Titanium postfilter 
allows to switch off the dominant 24 keV beam component and 
with it the neutrino-like recoils, leaving the scarce backgrounds intact \protect\cite{other}. 
To illustrate its effect, the vertical arrows 
mark the energy at which the endpoint of these soft recoils is 
predicted, 
based on a full MCNP-Polimi simulation \protect\cite{polimi} of the experiment and the 
expected 20\% quenching factor. {\it Inset:}  
Signals time-coincident with the thermal peak of a large 
$^{6}$LiI[Eu] scintillator mounted on a goniometric table allow to 
select discrete Ge recoil energies, for which quenching factors can then 
be obtained \protect\cite{other}. An excellent agreement with Lindhard theory expectations 
has been observed over the range of recoil 
energies relevant to the upcoming reactor neutrino experiment (see 
text).}
\end{figure}

\begin{figure}
\includegraphics[width=9cm]{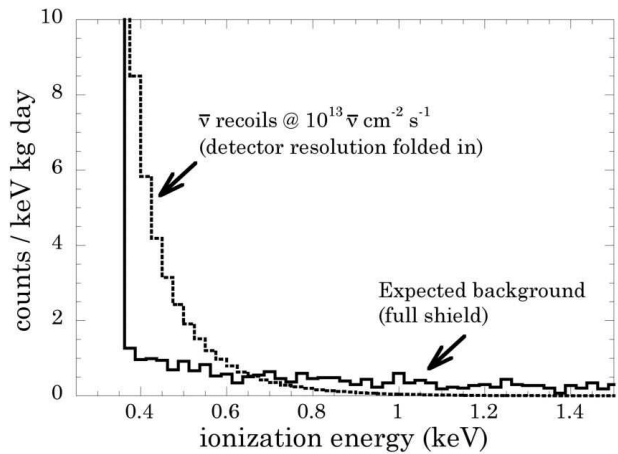}
\caption{\label{fig:epsart} Expected antineutrino signal in the planned 
demonstration
experiment, clearly visible above the background goal. The background 
is scaled down from data acquired with a partial 
shielding in place, i.e., the spectral shape depicted is representative of 
actual observations.}
\end{figure}

\begin{figure}
\includegraphics[width=9cm]{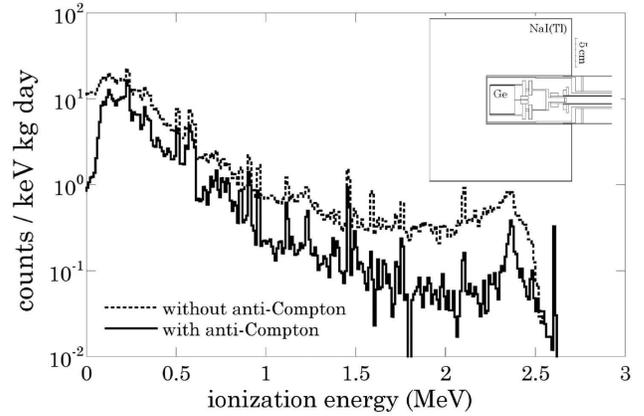}
\caption{\label{fig:epsart}Simulated effect of the anti-Compton veto in reducing the 
dominant contribution from radiocontamination in the preamplifier components 
($\sim$4 Bq $^{238}$U, $\sim$6.5 Bq $^{232}$Th, $\sim$0.2 Bq $^{40}$K).}
\end{figure}

\begin{figure}
\includegraphics[width=9cm]{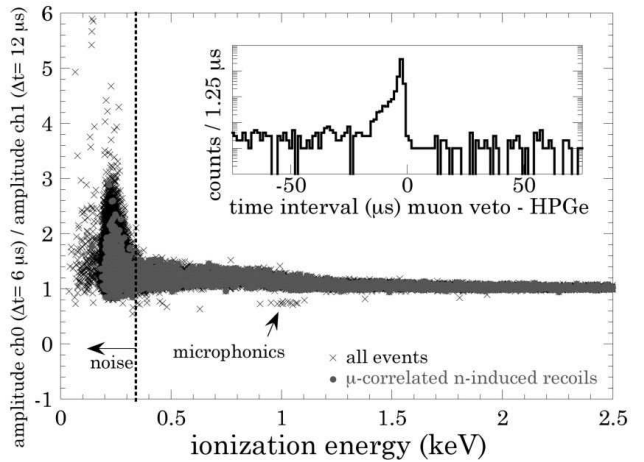}
\caption{\label{fig:epsart}Application of an analog method to reject  
anomalous pulses such as microphonics \protect \cite{morales} in the energy region near the 
detector threshold. Low-energy nuclear recoils produced by veto-tagged 
muon-induced neutrons (inset), characteristically delayed by the few 
$\mu$s of neutron straggling, are used to generate a template of "good" 
low-energy events (solid dots). Only a scarce number of microphonic events 
and a small fraction of 
electronic noise below threshold are observed to clearly deviate from 
it.}
\end{figure}

Motivated by this potential for progress and aiming at obtaining detector specifications 
sufficient for a demonstration measurement of the coherent 
neutrino-nucleus scattering cross-section, a modified electrode 
HPGe detector was built by CANBERRA Industries during 2005. 
Besides the mentioned improvements to the electronics, another 
departure from the approach in \cite{luke2} was the choice of p-type 
rather than n-type electrode configuration. The first reason for this 
was to enjoy 
a diminished sensitivity to very low energy minimum ionizing 
backgrounds (surface betas, x-rays, etc.), profiting from 
the $\sim$0.5 mm deep dead layer imposed by the n+ lithium-drifted contact 
that spans most of the surface of a p-type HPGe. The second was to 
depart from the observed strong degradation in the energy resolution 
of the detector in \cite{luke2}, which can be ascribed to electron 
trapping in dislocations as they travel relatively long distances 
through the crystal to 
the n+ central electrode in that geometry. It is for this reason that 
p-type detectors generally exhibit better 
charge collection than n-type, and this was deemed especially
important given the poor drift fields encountered with a modified electrode geometry.

The prototype displays an energy 
resolution (measured with a pulser) of 140 eV FWHM and a charge 
collection (\frenchspacing{Fig. 1}) comparable to that from 
a conventional coaxial-electrode HPGe. To put its performance 
in perspective, besides a much reduced noise level and 
threshold, it shows an 
enhancement in energy resolution from a much degraded value of $\sim$15 FWHM 
($^{60}$Co, 1.33 MeV) 
measured in \cite{luke2} to a 1.82 KeV FWHM, typical of a modern large 
coaxial 
HPGe. Charge collection uniformity along the axis varies by less than 0.15\% 
(\frenchspacing{Fig. 1}), compared to  3\% in \cite{luke2}. In 
other words, the device has the mass (475 g) and energy resolution typical 
of a large HPGe gamma spectrometer while simultaneously displaying 
the low-noise and low energy threshold (\frenchspacing{Fig. 2}) of an
x-ray detector a hundred times lighter. This unique combination makes the device a first of 
a kind, opening up a number of other interesting applications 
discussed in the next section. It is 
the first time that such a combination of mass, 
energy threshold and resolution has been 
produced in a radiation detector of any kind. 

\begin{figure}
\includegraphics[width=9cm]{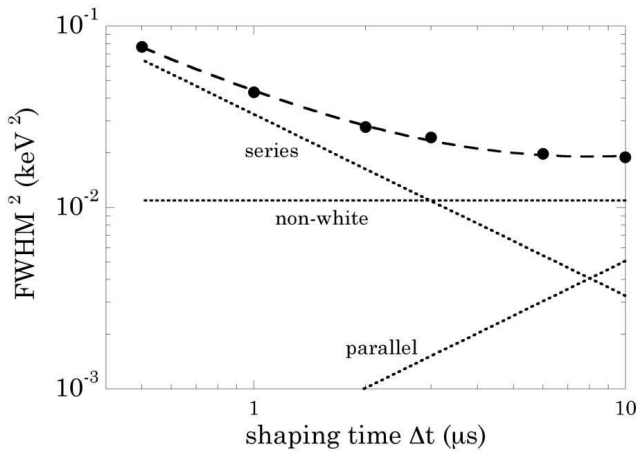}
\caption{\label{fig:epsart}Measured electronic noise components in the  
prototype (see text).}
\end{figure}

From the point of view of target mass and sensitivity to very 
small energy depositions the
device is ready for a  measurement of coherent neutrino scattering
in a reactor experiment: the inset in \frenchspacing{Fig. 3} 
displays the fraction of recoils 
above a choice of three different ionization energy thresholds E$_{thr}$ (i.e., 
the fraction of detectable events) as a 
function of their deposited (ionization) energy. The dotted line 
corresponds to pulser measurements (black dots) using this prototype 
(E$_{thr}=$ 330 eV). To make these measurements possible, 
a careful energy calibration of the pulser was 
performed with the help of five  gamma and x-ray lines 
identifiable in shielded conditions over the range 10-100 keV. 
The dashed line corresponds to envisioned minimal 
upgrades to the JFET (see below), resulting in E$_{thr}=$ 200 eV. The dash-dotted line 
illustrates the ultimate performance (E$_{thr}\sim$ 100 eV) that can be 
expected with present-day JFET technology, matching the state-of-the-art 
in low-noise x-ray detectors \cite{outdate,pentafet}. The 
combined effect of threshold 
and energy resolution can then be folded into the expected 
differential recoil rate in a reactor 
experiment (main plot, \frenchspacing{Fig. 3}), leading to 
integrated detectable coherent scattering rates of $\sim$2.5, $\sim$6 and $\sim$28 events / 
kg day, respectively. 

The quenching factor used for these calculations is 
the standard Lindhard theory prediction for Ge (k=0.2 in the notation 
of \cite{ge}). 
Measurements of this quenching 
factor in the interval of recoil energies 0.3-1.4 keV have been recently 
performed at the dedicated neutron beam described in a companion 
paper \cite{other}, using the existing prototype. While the many details of 
the analysis must be necessarily  
reserved for another publication \cite{new2}, \frenchspacing{Fig. 4}
illustrates the main results from this essential calibration, 
showing an excellent agreement with Lindhard-based expectations. It must be 
emphasized that the same range of recoil energies relevant to 
the upcoming reactor experiment (\frenchspacing{Fig. 5}) has been 
explored in this calibration (\frenchspacing{Fig. 4}), a crucial 
requirement before embarking in such a delicate measurement. To our 
knowledge, it is 
the first time that recoils this low in energy are directly and 
individually recorded 
(the measurements in \cite{ge} are indirect, 
i.e., based on the sum 
of recoil energy and a 69 keV nuclear de-excitation gamma).

Statistical evidence for coherent scattering can be gathered with the 
present prototype within just a 
few months of data acquisition near a power reactor, once the 
background goal depicted in \frenchspacing{Fig. 5} is reached. The 
disappearance of the neutrino signal during the refueling of the 
reactor (spanning 1-3 mo. every 12-18 mo.) would confirm the 
measurement. Shielded runs at a depth of 6 \frenchspacing{m.w.e.} and
extensive MCNP4b simulations \cite{mcnp} of U,Th, and 
$^{40}$K 
gamma and beta emissions in preamplifier parts, crystal holder and 
endcap indicate that the projected level of 
background is reachable with adequate 
active and passive gamma and neutron shielding in combination with a NaI(Tl) anti-Compton shield 
(\frenchspacing{Fig. 6}).
At least two precedents exist \cite{prec} for achieving 
O(1) count/keV kg day in the few keV region while benefiting from just a shallow 
(few \frenchspacing{m.w.e.}) overburden, what is expected at best in a 
reactor site (e.g., in a ``tendon'' gallery $\sim$20 m 
from the core \cite{adam2}). 
An important source of non-radioactive background in 
the low energy 
region is microphonic noise: pulse shape 
discrimination (PSD) techniques (analog \cite{morales}
and digital \cite{klapdor})
can be applied against it. With the single precautions of  
vibration-absorbing pads under the Dewar and attention to mechanically decoupling the 
cryostat from shielding materials, the prototype is observed to 
be rather insensitive to these (\frenchspacing{Fig. 7}). This is possibly a result of the 
very small capacitances involved \cite{microph}. 

It is worth 
emphasizing that contrary to what is implied in \cite{texono}, PSD 
techniques alone cannot be used to reduce an energy threshold 
imposed by electronic noise, 
unless an inadmissible penalty in signal acceptance is paid. In other 
words, any PSD cuts must be justified on grounds of clear physical 
differences between signal and noise pulses, as in \cite{morales}. 
\frenchspacing{Fig. 7} illustrates the indistinguishability of most 
true signals and noise below threshold.
Concentration should be in reducing electronic noise at the hardware 
level. 
Towards this end, the noise components in this prototype have been 
characterized.
As evidenced in the discussion of \frenchspacing{Fig. 3} the technique has not yet 
met the limitations imposed by present-day JFET technology: a threshold 
lower by a factor of $\sim\!3$ should be presently feasible with further 
effort, leading to a $>\!10$ 
enhancement in the signal rate per kg of detector mass for this neutrino process. An 
analysis of the electronic noise components in the existing prototype 
 (\frenchspacing{Fig. 8}) indicates that while the expectations 
described above for the series component  have been met 
(i.e., noise determined by the JFET and detector capacitances, \frenchspacing{Eq. 1}), a 
non-white component presently dominates. The origin of this is generally 
traceable to lossy dielectrics. Rearrangement and selection of the JFET 
package components should be able to lower this component to more 
manageable levels \cite{jf} and to generate an electronic noise behavior truly 
comparable to that of the best $\sim1$ pF x-ray detectors presently available. A 
small decrease in leakage current would also benefit the parallel 
noise component, leading to the (presently) ultimate performance,
i.e., few tens of detectable recoils per kg of Ge per day in the vicinity 
($\sim$ 20 m) of a high-power 
reactor. We note that up to an additional order of magnitude in signal rate 
could be obtained from a dedicated effort to reduce noise levels 
beyond the present state-of-the-art. 

While the detection of this intriguing mode of neutrino interaction seems finally within 
reach, many applications, either technological or in the study of 
fundamental neutrino properties
would need to involve  small arrays ($\sim$10 elements) of these detectors: such 
HPGe clusters are  common nowadays \cite{array}. 
Recent advances in HPGe detector encapsulation \cite{encaps} (leading 
to simplified handling and enhanced stability of array elements) and 
microphonic-free
cryocooling of HPGe arrays \cite{cool} can be applied to the 
construction of truly {\it compact} coherent neutrino detectors comprised 
of a modest number of 
modified-electrode crystals. In the specific realm of reactor 
monitoring applications, their sensitivity (hundreds of events per 
day) would then rival 
more cumbersome approaches based 
on underground ton-sized scintillator tanks \cite{bernstein1,adam2}. 

\section{APPLICATIONS BEYOND COHERENT NEUTRINO SCATTERING}

The interest of a large-mass, ultra-low threshold semiconductor 
detector is not limited to coherent neutrino detection. Experimental 
searches for Weakly Interacting Massive Particles (WIMPs) \cite{revexp}, prime dark 
matter particle
candidates, typically assume that these particles are gravitationally 
bound to our galaxy, resulting in expected speeds relative to Earth of 
O(300) km/s. This in turn determines the recoil 
energies they are to deposit in dedicated detectors (few keV to tens 
of keV, depending on their mass). An alternative WIMP population 
consists of those that fell into stable closed solar orbits during the 
formation of the protosolar nebula, following colissionless 
mechanisms \cite{steigman}. In that case their maximum speed is limited 
by the escape velocity from the Sun at Earth (42 km/s), leading to a 
large
concentration of the recoil signal below the threshold of present-day WIMP 
detectors \cite{solar} but now within reach of this new device. A second assumption 
driving most WIMP searches is the false tenet that the only supersymmetric 
WIMP candidate is the so-called neutralino \cite{revexp}: as has  
been recently emphasized \cite{kuse}, a parallel phenomenology exists 
for other non-pointlike supersymmetric dark matter candidates (e.g., Q-balls), 
one that 
involves very soft recoils, again generally below threshold and hence 
beyond detection for more conventional detector designs. 
A third application in this field of dark matter 
detection is the exploration of the possibility 
that a light (5-12 GeV/$c^{2}$) particle might be responsible for the
controversial WIMP DAMA signal \cite{dama}, a hypothesis not 
contradicted 
yet by any other experiments \cite{sd,si}.
The first
experimental limits on these interesting possibilities can be 
extracted from background characterization runs for the existing 
prototype \cite{new2}, with a sensitivity that should increase as the 
technique is further developed. Ironically, 
WIMP interaction rates like those in \frenchspacing{Fig. 9} would 
frustrate 
any attempt to detect reactor neutrinos coherently. Separately from the subject of dark 
matter searches, it is also worth remembering that the present sensitivity to 
a finite neutrino magnetic moment should be increased by about an order of 
magnitude, down to $\mu_{\nu}(\bar{\nu}_{e})\rightarrow \sim2\times 10^{-11} 
\mu_{B}$ in a detector with these characteristics \cite{wmm}.

\begin{figure}
\includegraphics[width=9cm]{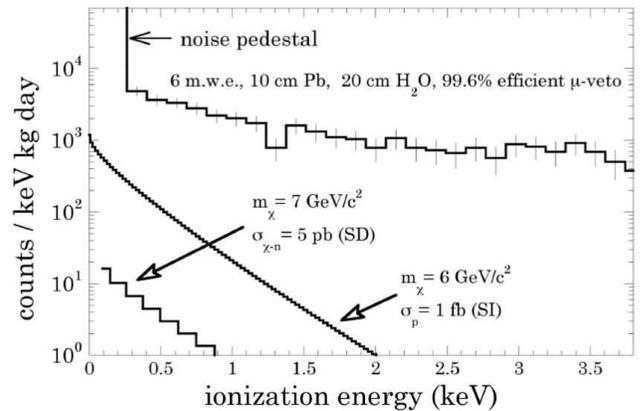}
\caption{\label{fig:epsart}Signals expected from typical WIMP 
candidates able to explain the DAMA annual modulation \protect \cite{dama} while satisfying the 
limits imposed by all other experiments \protect \cite{sd,si}. The 
differential spectra are labelled by WIMP mass and proton 
cross-section (SI$=$ spin-independent coupling, 
SD$=$ spin-dependent).
The prototype data were acquired with  
a partial shield. The addition of a NaI 
anti-Compton veto (expected $\times$10 reduction in background), 
replacement of cryostat endcap and crystal holder with low-activity Al ($\times$50 
reduction) and 
additional Pb and overburden should allow to confirm or rule out 
these light dark 
matter candidates.}
\end{figure}

\begin{figure}
\includegraphics[width=8.7cm]{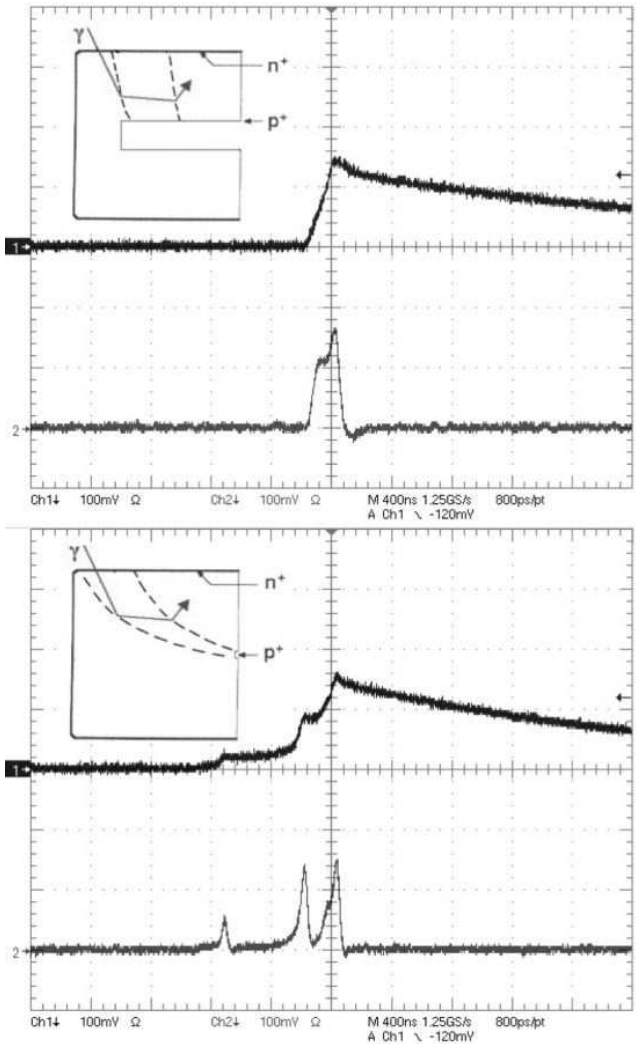}
\caption{\label{fig:epsart}Effect of electrode geometry on pulse formation 
for a multiple-site gamma interaction (see text).}
\end{figure}

Another important application of a modified electrode HPGe arises 
from exploiting the characteristic temporal features of the preamplifier 
traces in this electrode configuration, 
features already noted in \cite{luke2}:
\frenchspacing{Fig. 10} displays two sets of traces, the top panel for 
a conventional coaxial electrode detector, the bottom panel for the prototype 
under discussion, both crystals roughly the same mass and dimensions. The top 
trace in each panel corresponds to the preamplifier output, the 
bottom one is this same passed through a Timing Filter Amplifier (TFA) set 
at 10 ns differentiation and 10 ns integration time. Such a TFA trace is a good 
representation of the time-structure of the arrival of charge to the 
collecting electrodes \cite{klapdor} (the same signal 
processing can be 
achieved on software using digitized preamplifier traces). 
For both detectors the event 
that originated the pulses was a multiple-site gamma 
interaction. In the coaxial TFA trace this is barely evident, a 
double ``hump'' or somewhat broadened trace being the only signature, 
resulting in a limited ability to differentiate multiple from 
single-site 
interactions \cite{klapdor}. In the present prototype however, the 
longer distances that charge (holes in this case) must travel to the 
small p+ electrode result in traces generally stretched by a factor 
$\sim$5 in time, yielding a clear separation of individual site 
 contributions to the overall pulse (a three, possibly four-site interaction in 
the figure). It is worth noting that the reduced electronic noise is in 
itself an advantage in this type of pulse shape analysis 
\cite{match}. The insets in \frenchspacing{Fig. 10} graphically illustrate how the radial 
degeneracy imposed by the coaxial electrode dissolves for the modified 
electrode: in the first case, the trajectories along the dashed field lines followed by 
electrons and holes generated 
at similar radial distances in the crystal are comparable, leading to 
a trace that would most probably be misidentified as corresponding to 
a single-site interaction. In the second, for the same initial distribution of charge, 
the hole trajectories are 
considerably different, leading to positive 
identification of the event as a multi-site interaction. In 
\frenchspacing{Fig. 11} this effect is illustrated  
with the help of a 59.5 keV $^{241}$Am collimated gamma source impinging on the 
crystal perpendicularly to its axis and, given the low gamma energy, interacting 
via single-site photoelectric effect very close to the 
surface. Little difference is observed as 
a function of axial source position in coaxial traces (top), whereas 
the prototype traces (bottom) are clearly rich in position information. In this 
respect and as a simple curiosity, 
keeping in mind that the charge drift speed in depleted Ge is 
$\sim$10 ns/mm, it is possible to extract the rough crystal 
radius (2.5 cm in actuality)
from the rise time of the d=43 mm trace (the closest gamma injection 
point to the p+ electrode, at the base of the crystal), 
as well as the approximate crystal length (4.4 cm) 
from the time dispersion in the onset 
of the traces.

\begin{figure}
\includegraphics[width=9cm]{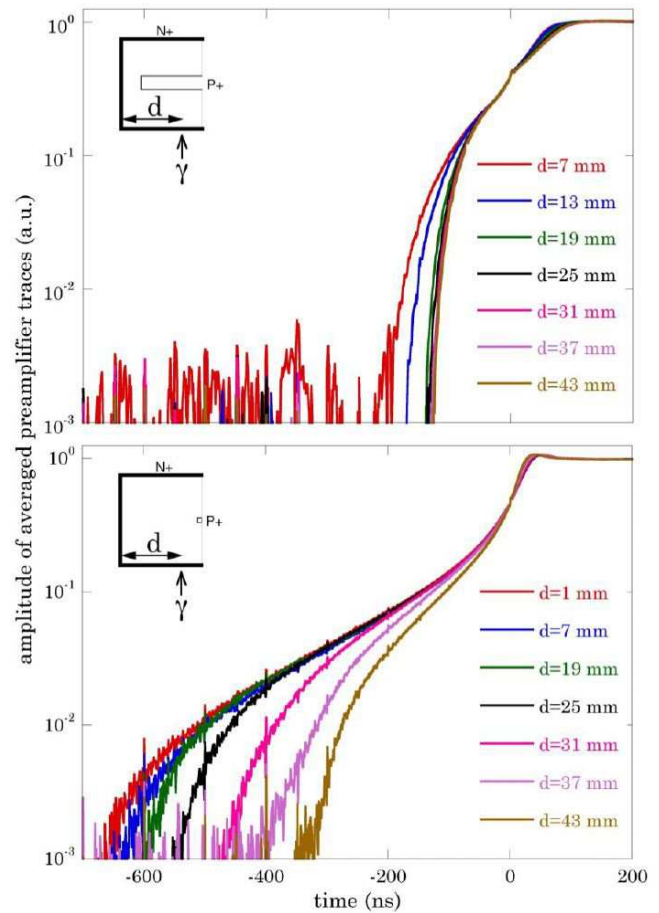}
\caption{\label{fig:epsart}Normalized preamplifier rising edges for a coaxial 
(top) and modified electrode (bottom) HPGe (see text).}
\end{figure}

\begin{figure}
\includegraphics[width=8.7cm]{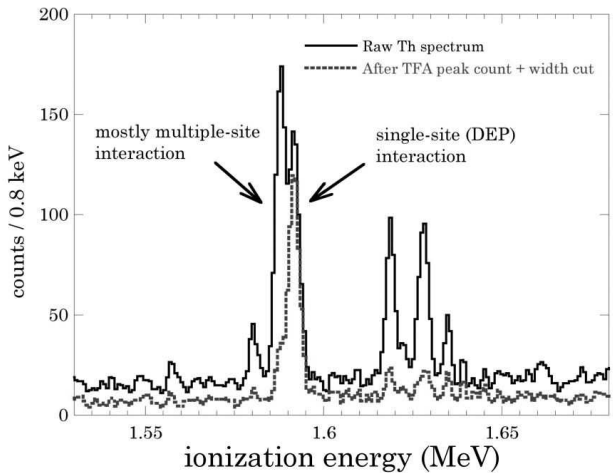}
\caption{\label{fig:epsart}Natural Thorium irradiation of the modified 
electrode p-type detector and effect of PSD cuts on rejection of 
multiple-site events (see text). The energy resolution appears slightly 
degraded due to the limited resolution of the digitizer employed (8 
bit).}
\end{figure}

Besides the evident commercial applications that distinguishing 
single from multiple-site interactions might have (e.g., an effective higher 
peak-to-Compton ratio for smallish crystals), next-generation neutrinoless 
double-beta decay searches using germanium crystals 
as source and target (MAJORANA \cite{maj}, GERDA \cite{ger}) rely 
heavily on this distinction: the process of interest, the local emission of 
two short-ranged $\sim$1 MeV electrons indicating a neutrinoless decay, 
is eminently single-site, 
while a large fraction of backgrounds is not. The mentioned limited 
performance in this respect from conventional coaxial crystals  has 
lead these experimental efforts to consider complex multi-electrode 
segmented germanium detectors, where the additional spatial 
granularity helps improve the background rejection ability. 
Evidently, the prohibitive energy resolution degradation observed in 
n-type modified electrode schemes \cite{luke2} would never allow one to 
consider a neutrinoless double-beta decay application, where the 
signature sought is a very precisely defined 
energy deposition corresponding to the Q-value of the nuclear 
transition (2039 keV for $^{76}$Ge). The optimal charge collection 
observed in a p-type design leads on the other hand to excellent 
prospects. 
\frenchspacing{Fig. 12} shows the results of a prototype irradiation 
with a natural Th source. Following \cite{th}, the $^{228}$Ac gamma ray at 1588 
keV typically interacts three times to deposit its full energy, 
whereas the line at 1592 keV arises from a double escape (i.e., a 
single interaction) from a 2614 keV emission in $^{208}$Tl. Their 
neighboring energies 
provide a convenient benchmark to asses the ability of a detector to distinguish 
single from multiple-site interactions \cite{th}. A simple,
energy-independent set of cuts based on peak counting of digitized TFA traces 
such as those in \frenchspacing{Fig. 10} leads to a background 
(multiple-site) rejection (BR) of 88\% for a signal acceptance (SA) of 93\%
(\frenchspacing{Fig. 12}). This compares very favorably, even prior 
to the beneficial effect of close-packing crystals, to an 8-channel 
segmented four-crystal ``clover'' HPGe detector (93\% BR for 73\% 
SA) \cite{steve}. In the case of the modified electrode prototype, application of 
the same cuts to the neutrinoless double beta decay region of 
interest (2.04 MeV) leads to a reduction in background by $\sim$50\%, in 
close agreement with the expectations from a simulation of this 
particular Th irradiation. In other words, the bulk of 
multiple-site backgrounds is 
successfully rejected.

Besides the obvious advantages that use of a single-channel device can bring 
to next generation germanium double-beta decay searches (increased 
speed of deployment, reduction in costs, simplified construction and 
analysis) there are other more subtle ones that can be listed: first, 
the
largest fraction of the backgrounds is expected to arise from front-end 
electronics and their cabling, evidently much more numerous in a 
segmented detector approach. 
Second, the thermal load imposed by the excess cabling involved in 
segmented schemes is much reduced. Third, 
long-term instabilities 
and channel cross-talk are known undesirable features in some 
segmented configurations. Fourth, segmentation schemes involve 
use of n-type crystals: leaving aside the slightly lesser energy resolution 
these usually display (however crucial in double-beta), p-type 
crystals are more rugged, which is advantageous when 
arraying. They should also profit from the thick lithium-drifted dead layer to 
exhibit a much lower sensitivity to surface contaminations, implanted 
and absorbed
alpha-emitting
Rn daughters being a major concern in double-beta decay experiments.

\section{CONCLUSIONS}

A novel type of radiation detector, a large-mass p-type modified electrode HPGe 
diode has been presented. It features an unprecedented combination of 
sensitivity to small energy depositions, excellent energy resolution, 
large mass and built-in information about site multiplicity of the 
interactions: as is often the case, a 
promising new addition to the existing arsenal of radiation detection technologies 
can generate a number of exciting applications. In this case, 
the device can be considered a first viable coherent neutrino 
detector, allowing to realistically consider the possible technological 
applications that smallish neutrino detectors may one day find. 
In addition to this, a strong role in double-beta decay, neutrino 
magnetic moment and dark matter 
searches can be foreseen.

\section{ACKNOWLEDGEMENTS}
This work is supported by NSF CAREER award PHY-0239812, DOE/NNSA 
grant DE-FG52-0-5NA25686, and in part by 
the Kavli Institute for Cosmological Physics through grant NSF 
PHY-0114422 and a Research Innovation Award 
No. RI0917 (Research Corporation). 
We are indebted to J. Lenhert and T. 
Wilson at NNSA/NA-22 for their resolved support of this project and to 
L. Darken and D. Gutknecht for their contributions to crystal growing 
and detector fabrication. 
Similarly, to C. Aalseth, F.T. Avignone, J. Beacom, R.L. Brodzinski, G. 
Gelmini, P. Gondolo, A. Kusenko, P. Luke, H.S. Miley, B. Odom, D. Radford, J. Wilkerson and 
the MAJORANA collaboration for many useful exchanges.  
This work is dedicated to the memory of R.L. Brodzinski.

\end{document}